\documentclass[mathleft
]{an}
\usepackage{graphicx}
\usepackage{times}
\usepackage{tabularx}
\begin{document}

\Pagespan{789}{}
\Yearpublication{2006}%
\Yearsubmission{2005}%
\Month{11}%
\Volume{999}%
\Issue{88}%

\title{The FIRST radio survey: Panchromatic properties of FIRST radio
  sources identified in  the Bo\"{o}tes and Cetus fields}

\author{K. El Bouchefry\inst{1}\fnmsep\thanks{
  \email{kelbouchefry@gmail.com}\newline}
}

\titlerunning{EROs counterparts to FIRST}
\authorrunning{ El Bouchefry  2007}
\institute{ Astrophysics and Cosmology Research Unit,  University of KwaZulu-Natal, Westville, 4000, South Africa}

\received{2008 Apr 30} \accepted{2008 Aug 19}
\publonline{2008 Dec 28}


\abstract{%
  In this paper, the second in a series,  the availability of multi-wavelength
  optical/infrared information of FIRST (Faint Images of the Radio Sky at 20
  cm) radio sources counterparts over $\sim$ 9.2 deg$^{2}$ in Bo\"{o}tes
  field and $\sim$ 2.4  deg$^{2}$ in Cetus field is exploited to infer the
  physical properties of the faint radio population. The radio sources
  optically  identified have been divided into   resolved galaxies and
  stellar-like objects finding that  the faint radio   population is mainly
  composed of early-type galaxies with  very red colour   ($Bw-R\sim 4.6$). A total number of $57$ counterparts of FIRST radio sources have   extremely red
  colour ($R-K\geq 5$). Photometric redshift from \textit{Hyperz} implies
  that the Extremely Red Objects (EROs) counterparts to FIRST radio sources are mostly located in the
  range $z=0.7-2 $, with the bulk of the population at $ z\sim 1$. Taking
  advantage of the near infrared imaging  with FLAMEX (FLAMINGOS Extragalactic
  Infrared Survey), the EROs counterparts to FIRST radio sources are separated  into passively-evolving and dusty
  star-forming galaxies using their \textit{RJK} colours; the relatively blue $J-K$  of
  these galaxies  suggest that most  are old elliptical galaxies (18/25)
  rather  than dusty starburst  ($7/25$). A total of  15 Distant Red   Galaxy (DRGs) have been identified as counterparts to  FIRST radio sources in Cetus field and 3   DRGs in Bo\"{o}tes field with  $J-K>2.3$.}   \maketitle

\section{Introduction}

The radio sources in deep radio surveys at 1.4 GHz (20 cm) consist of two main
populations; active galactic nuclei and star forming galaxies (Condon  1984,
Windhorst et al. 1985). It has long been known that powerful radio sources are
associated with AGN or Giant elliptical (e.g. Windhorst 1990). At mJy level
and fainter,  deep radio surveys  at 1.4 GHz (Condon 1984, Windhorst et
al. 1985. Condon 1992) at mJy level and fainter have shown that the faint
radio population is a mixture of different classes of objects (radio loud, and
radio quite AGN, starburst galaxies, spirals). To date  the
nature and the level of contribution  of each population is still not well established, and very little is known about the  cosmological evolution of different
kinds of objects and so many other topics despite so many studies  (e.g. Georgakakis et al. 1999;
Gruppioni et al. 1999;  Ciliegi et al. 2003; Hopkins et al. 2003;  Prandoni et al. 2006; Fomalont et al.  2006; Simpson et
al. 2006;  Bondi et al. 2007).  This is mainly due to  the
incompleteness of optical identification and the optical depth of the
spectroscopic follow up. Therefore, large and statistically complete and
homogeneously selected samples, very deep optical imaging and spectroscopic follow up
are required for reasonably large deep radio samples in order to investigate
the nature and evolution of the faint radio population.  

Studies of Extragalactic radio sources are among the most interesting
challenges of modern cosmology and astrophysics. These studies have been invigorated due to the recent 
generation of panchromatic photometric and spectroscopic large area surveys,
such as  NVSS (Condon et al. 1998), FIRST (Becker et al. 1995), SDSS (York
et al 2000), and 2DF (Colless et al. 2001). These surveys have provided additional
panchromatic photometric and optical/infrared spectroscopic observations
(Sadler et al. 1999, Best et al. 2005). For example, Ivezic et al. (2002), Best
et al. (2005) and Obric et al. (2006) cross correlated FIRST and SDSS, and
analysed the optical and radio properties of quasars and galaxies in full
details.  Best (2004) studied the environmental dependence of radio luminous
AGN and Star Forming galaxies, and their luminosity function has been investigated in
other studies (Sadler et al. 1999, Jackson \& Lindish 2000, Chan et al. 2004,
Best et al. 2005). Recent searches have been undertaken to search for
clustering of galaxies (Hall et al. 2001) and extremely red objects (EROs)
around high redshift quasars and radio galaxies (Cimatti et al. 2000, Wold et
al. 2003, Zheng et al. 2005, and references therein). EROs are  of special
interest in the study of galaxy evolution, in that their colours and other
properties suggest that  they are the high-redshift ($z=1-2$) counterparts and
progenitor of local elliptical and So galaxies, and are amongst the oldest
galaxies present at these redshifts.

The present paper is the second in a series analysing the properties of the
FIRST radio sources which are identified in NDWFS/FLAMEX surveys. In paper I,
the author has identified FIRST radio sources in Bo\"{o}tes field over a  region of
$\sim$ 9.2 deg$^{2}$ and in Cetus field over a region of $\sim$
4.2 deg$^{2}$. The identification rate was found to be $63\%$ in Bw band,
$65\%$ in R band, $64\%$ in I band, $38\%$ in K band and  $39\%$ in four
bands (\textit{Bw, R, I} and \textit{ K}). I also  derived photometric redshifts of FIRST radio sources
counterparts  using the public code \textit{Hyperz}. In this paper (II) I want
to shed  some light into the optical/infrared properties of the
FIRST-Bo\"{o}tes/Cetus sample and their environment.

The layout of this paper is as follows:  Section 2, presents the 
magnitude-flux distribution of the FIRST counterparts in Bo\"{o}tes and Cetus
field. Section 3 discusses the properties of FIRST radio sources in Cetus field. Section 4 investigates the EROs counterparts to FIRST radio sources and section 5 summarises conclusions.

Throughout this paper it is assumed that H$_{\circ}$$=$70 km.s$^{-1}$
Mpc$^{-1}$, $\Omega_{M} =0.3$, and $\Omega_{\Lambda} = 0.7$ unless
stated otherwise.

\begin{figure}
  \begin{center}
    \begin{tabular}{c}
      \resizebox{77mm}{!}{\includegraphics{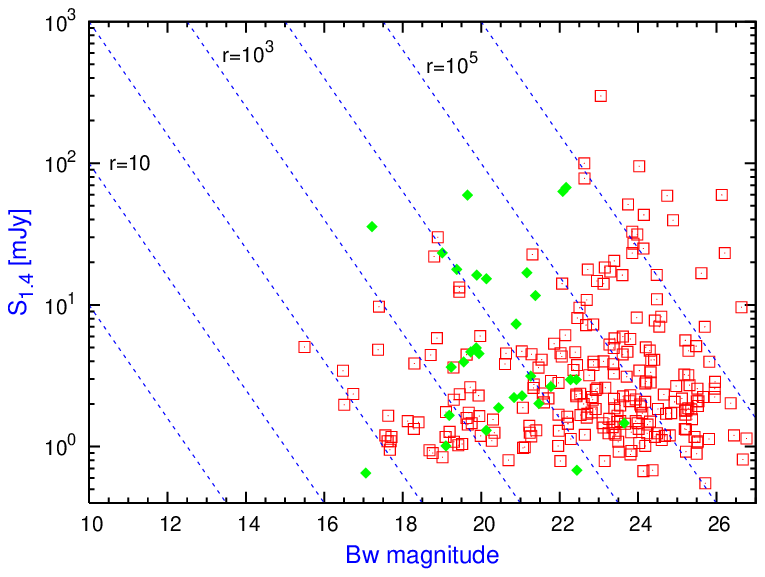} }\\
    \resizebox{77mm}{!}{\includegraphics{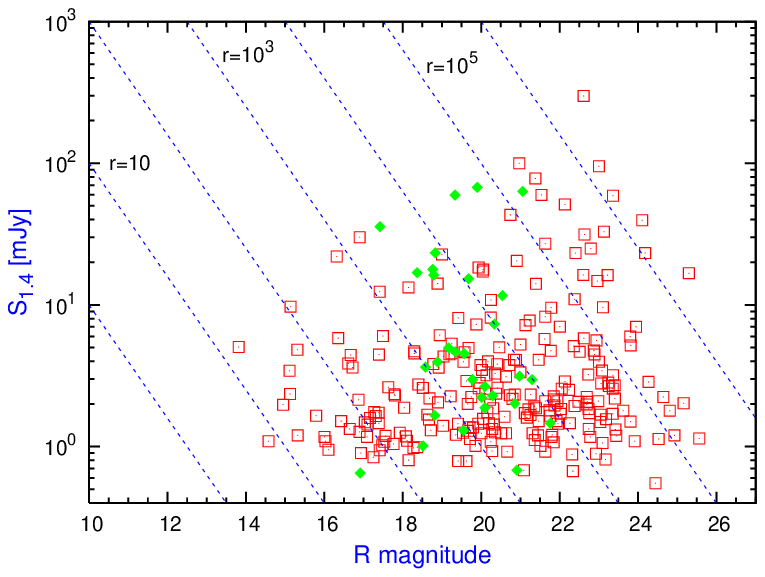} }\\
     \resizebox{77mm}{!}{\includegraphics{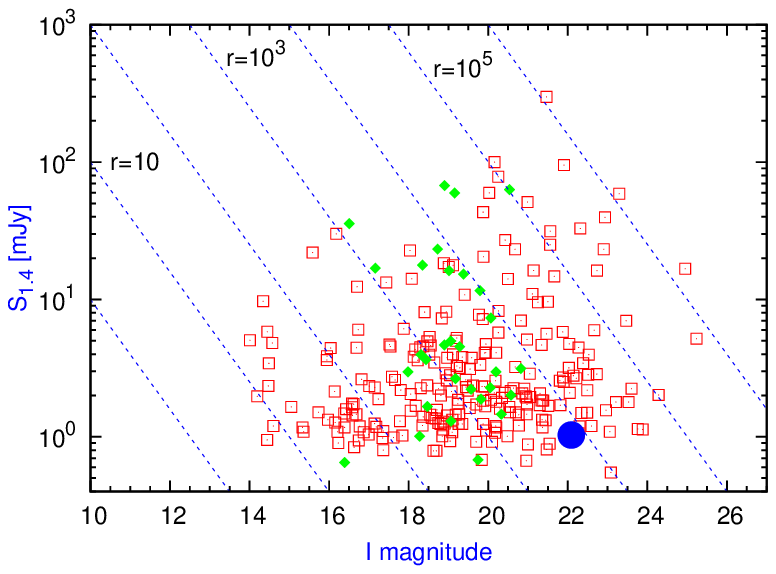} }\\
    \resizebox{77mm}{!}{\includegraphics{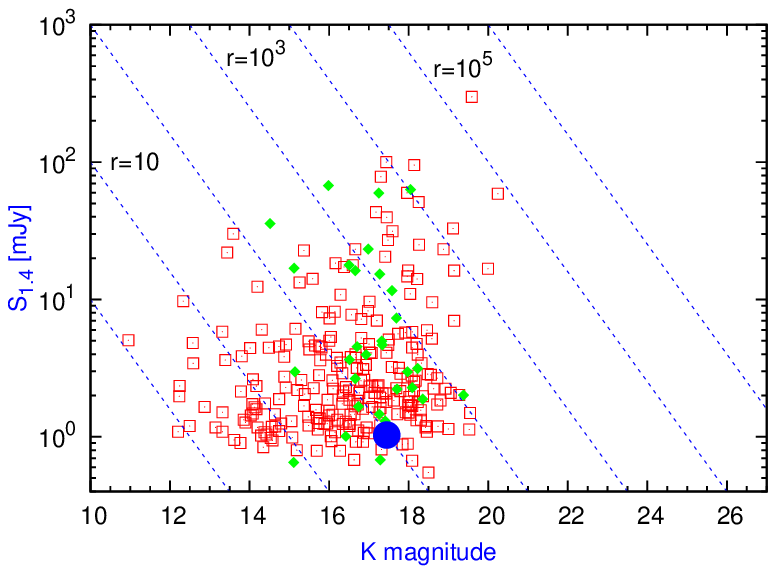} }
    \end{tabular}
   \caption{The radio flux vs. \textit{Bw, R, I} and \textit{K} magnitudes for
   all FIRST radio sources identified in Bo\"{o}tes field.
    Empty squares represent sources identified as  galaxies,
     filled diamonds indicates sources identified as  stellar-like
     objects and the large filled circle (in I and K bands)
      represents a radio-loud quasar (McGreer et al. 2006).
      Superimposed are the lines corresponding to constant
       values of the radio-to-optical ratio $r= 10,100, 10^3,10^4...10^6$} 
\label{magflux}
  \end{center}
\end{figure}
\section{Magnitude - Flux distribution in Bo\"{o}tes field}

The combination of the sample introduced in El Bouchefry \& Cress 2007 (hereafter EC07) and those introduced in paper I (El Bouchefry 2008)  provided  a total of 338 FIRST radio sources with
infrared counterparts and a total of 273 sources with identification in four
bands (\textit{Bw, R, I, K}). The FIRST radio sources  with optical
identifications have been divided into galaxies and stellar objects by means of the SExtractor
Class-Star parameter S/G (Bertin \& Arnouts 1995). This parameter is provided
by the NDWFS catalogue and ranges from 0.0 to 1.0 being the most point
like. Of the 273 IDs, 243 sources were identified as galaxies and 30 sources
identified as stellar objects. Combining the analysis of the radio and photometric
properties can provide a first indication of  the nature of the faint radio
population. Figure \ref{magflux} shows the 20 cm radio flux versus B, R, I and
K magnitudes for all FIRST radio sources with an optical/infrared
identifications. Superimposed are the lines corresponding to constant values
of the observed radio-to-optical ratio \textit{r}, defined by Condon (1980) as follows:

\begin{eqnarray}
r=S\,\times\,10^{\, 0.4\;(mag-12.5)}
\end{eqnarray}

\noindent where $S$ is the 1.4 GHz flux in mJy and mag is the apparent
magnitude of the optical/infrared counterparts. \\ 
In all diagrams empty squares stand for resolved sources (galaxies), filled diamonds
for unresolved sources (stellar objects) and filled circles represent a
radio-loud quasar (J142738.5+3312) discovered by McGreer et al. (2006) as a
counterpart to FIRST radio sources by combining NDWFS and FLAMEX data. A
radio-loud source is considered to have $r>10$ (Urry \& Padovani
1995). Magliocchetti, Celotti \& Danese (2002) showed that for star-burst
galaxies, $r_B<100$. Therefore, the \textit{Bw, R, I, K} magnitude flux diagrams
(Figure \ref{magflux}) show that there are a few star-forming galaxies in the
FIRST-NDWFS/FLAMEX sample down to 1 mJy. This result may indicate that
star-forming galaxies actually dominate the sub-mJy population. This is
also reported by Gruppioni et al. 1999b, in the study of the Marano field,
Georgakakis et al. (2005) in the Phoenix survey and Magliocchetti et
al. (2002) in the FIRST-APM sample.

\begin{figure}
\begin{center}
 \includegraphics[width=83mm]{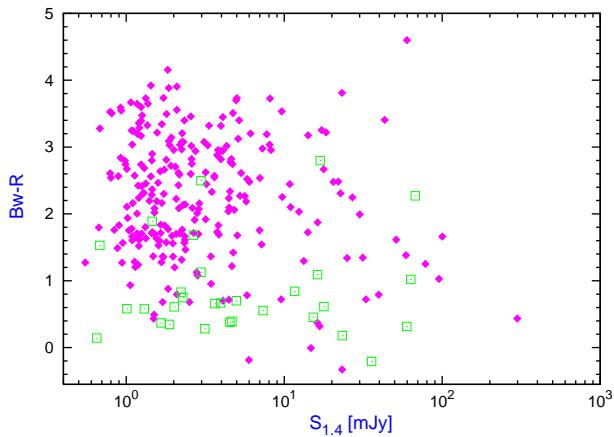}
\caption{The Colour Bw-R versus radio flux for all FIRST radio sources
  counterparts. Filled diamonds indicate sources identified as  galaxies
  while empty squares represent stellar-like objects} \label{brcolor}
\end{center}
\end{figure}

It is important to note how the radio population of radio galaxies is shifting
towards brighter red magnitudes (see Figure \ref{magflux}). The shift is
observed especially for resolved sources (galaxies) and not for the stellar
identifications, highlighting the fact that faint radio galaxies are typically
red, early-type objects. Figure \ref{brcolor} represents the
\textit{Bw-R} colour as a function of radio flux S$_{1.4}$ for the two populations: resolved
sources (filled diamonds) and stellar objects (empty squares). It clearly seen see
from the plot that, the majority of FIRST radio sources counterparts consists of
galaxies with very red colours, up to \textit{Bw-R} $\sim 4.6$. One also notes that, the radio
galaxies appear mostly with \textit{Bw - R }$>1$, and stellar-like objects dominate the region
$ 0<$ \textit{Bw -R} $< 1 $ leaving a very small fraction of  objects with blue colour
(\textit{Bw -R}$<0$) characteristic of star forming galaxies. Again this is a further
confirmation of the results of Magliocchetti \& Maddox (2002) and
Magliocchetti et al. (2004) in the radio-optical study of the FIRST radio
sources in APM and 2DF surveys respectively.

\section{Properties of FIRST radio sources in Cetus field}

Cross correlating the FIRST radio sources with sources in Cetus
field yielded 113/242 counterparts in \textit{J} band; 124 radio sources
identified in \textit{K} band and 109 sources identified in two bands (\textit{J,K}, see El Bouchefry 2008).

\begin{figure}
\begin{center}
 \includegraphics[width=83mm]{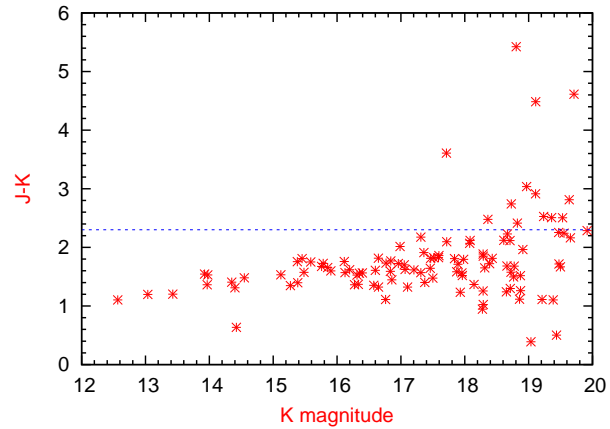}
\caption{The Colour magnitude sequence for  all FIRST radio sources
  identified in Cetus field.} \label{cetuscolor}
\end{center}
\end{figure}

\begin{figure}
\begin{center}
 \includegraphics[width=83mm]{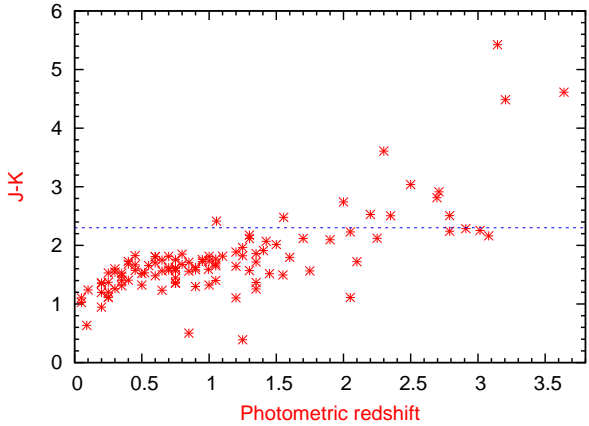}
\caption{The Colour $J-K$ against photometric redshift for  all FIRST radio sources
  identified in Cetus field.} \label{cetuscolorz}
\end{center}
\end{figure}

 Figures \ref{cetuscolor} and \ref{cetuscolorz}  illustrate the
\textit{J-K} colour against $K$ magnitude and photometric redshift
respectively. A total of  $15$ counterparts to FIRST radio sources in
this field have a red colour \textit{J-K} $> 2.3 $ that selects the
so-called Distant Red Galaxies (DRGs). This latter in turn select
an heterogeneous samples of galaxies going from the passively
evolving systems to a significant fraction of obscured star
forming galaxies at $2<z<4.5$ with strong Balmer or 4000
$A^{\circ}$ breaks. As a comparison, Hall et al. (2001) select
four galaxies with $J-K>2.5$ and found photometric redshift
$z\geq2$, Franx et al. (2003) found 14 DRGs with $J-K>2.3$ in a
photometric redshift range from $1.92$ to $4.26$ (see also
Takagai et al. 2007, Papovich et al. 2006).

\begin{figure}
\begin{center}
 \includegraphics[width=83mm]{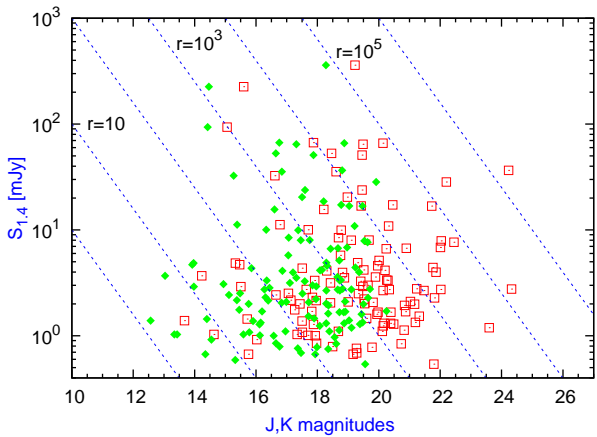}
\caption{The radio flux vs. the magnitude for FIRST radio sources
identified in Cetus field , in \textit{J} band (empty squares) and \textit{K} band
(filled diamonds). Dashed lines correspond  to constant values of
the radio-to-optical ratio $R=10,100,10^3,10^4...10^6$   }
\label{cetusmagflux}
\end{center}
\end{figure}

In the FIRST - Cetus sample, there are 3/15 DRGs counterparts to
FIRST radio sources at lower redshift ($1\leq z \leq 2$)
, and $8/15$ with photometric redshift ranging from $2$ to $3.8$ and
$17<K<20$. Papovich et al. (2006)  reported that lower redshift objects are
dominated by dusty star bursts and higher redshift are objects with more
complex stellar population, these are  likely to be passively evolving stellar components.
Obviously spectroscopic confirmation is required before  one
can draw any conclusion. Figure \ref{cetusmagflux} illustrates the
magnitude - flux diagram  for the  FIRST radio sources identified
in J (empty squares) and K (filled diamonds) bands. It is
noted again an interesting shift of the population towards brighter
magnitude with a small number of sources that have a radio-to-optical ratio in \textit{J} band $r\leq\,100$.

\section{Extremely red objects  counterparts to FIRST radio sources in Bo\"{o}tes field} 
\label{ero}
Extremely red objects were first discovered more than 20 years ago
(Elston et al. 1988) as resolved galaxies with $ K\sim 16.5$ and
\textit{R-K} $\sim5$. Photometry and spectroscopy showed that these objects were old
elliptical galaxies at $z=0.8$ (Elston et al. 1989). EROs seem to be found
everywhere; they were found in the vicinity of high redshift (
McCarthy et al. (1992); in a quasar field (Stockton et al. 2006;
Hu \& Ridgway 1994),  as counterparts of faint X-ray (Newsam et
al. 1997) and as a counterparts of radio sources (Smail et al.
2002; Willott, Rawlings \& Blundell  2001, De Breuck et al. 2001,
Spinrad et al. 1997). Their colours are consistent with two
heterogenous mix of galaxy classes: 1) Old, passively evolving
elliptical galaxies at $ z \geq 0.9$, the colour is due to the lack
of star formation and the large K-Correction, 2) dust reddened
star-forming galaxies or AGN (see e.g. Cimatti et al. 2003; Wold
et al. 2003; Yan \& Thompson 2003; Daddi et al. 2002; Smail et al. 2002;
Roche et al. 2002). The identification of either type offers a
potentially important deep insight into the formation and
evolution of elliptical galaxies and help to investigate the
existence of a population of dusty galaxies of AGN strongly
reddened by dust extinction.

\begin{figure*}
\begin{center}
 \begin{tabular}[!h]{cc}
 \resizebox{80mm}{!}{\includegraphics{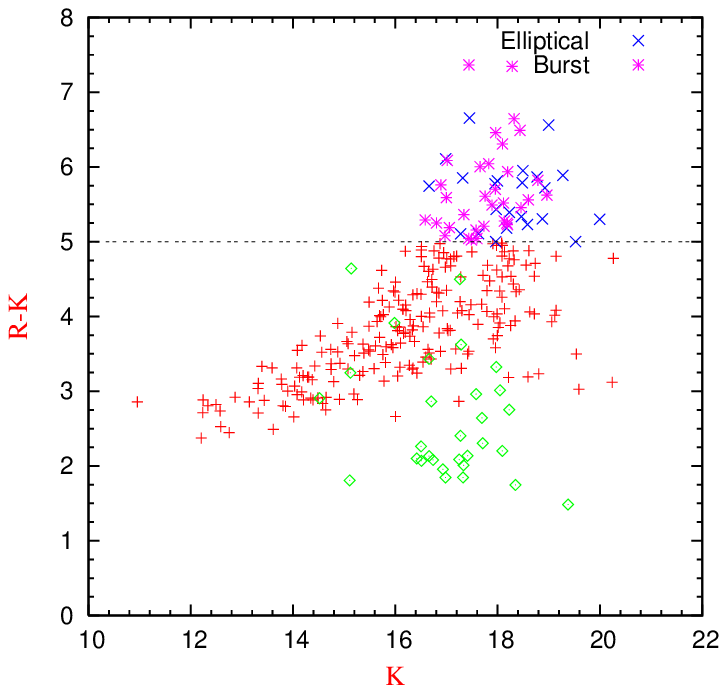} }
     \resizebox{80mm}{!}{\includegraphics{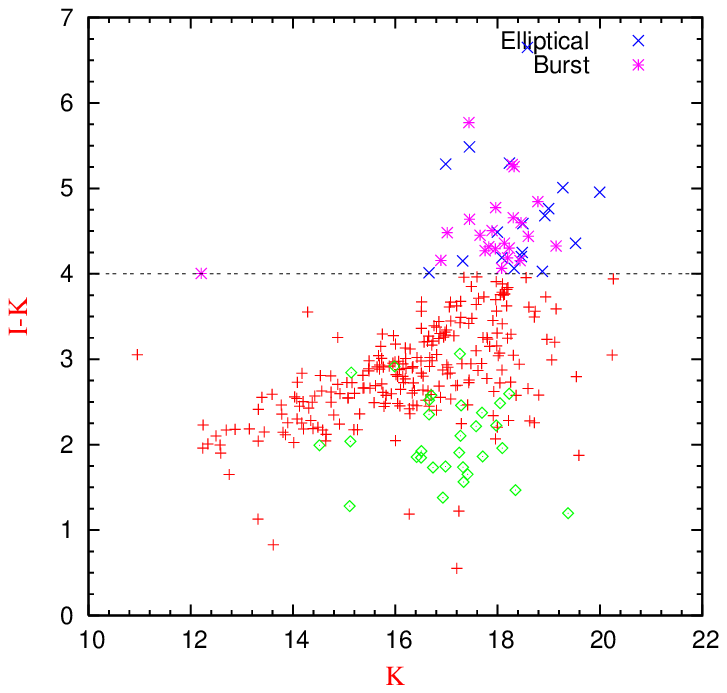} }
 \end{tabular}
\caption{The K band magnitude vs. the $R-K$ and $I-K$ colours, in
left and right panels, respectively. Empty diamonds indicate
stellar-like sources, while the crosses, stars and plus signs represent
objects identified as galaxies. The horizontal long-dashed line
corresponds to the limit adopted for the selection of the sample
of EROs counterparts to FIRST radio sources in Bo\"{o}tes field.}
\label{rkcolor}
\end{center}
\end{figure*}

\begin{figure*}
  \begin{center}
    \begin{tabular}[!h]{cc}
      \resizebox{83mm}{!}{\includegraphics{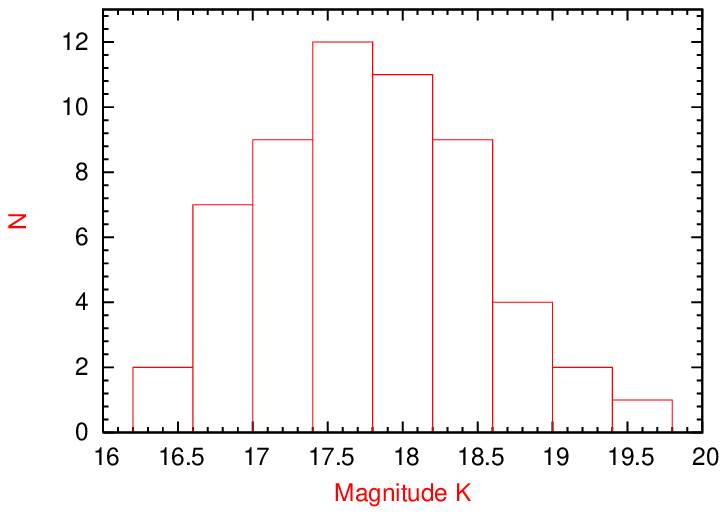} }
      \resizebox{83mm}{!}{\includegraphics{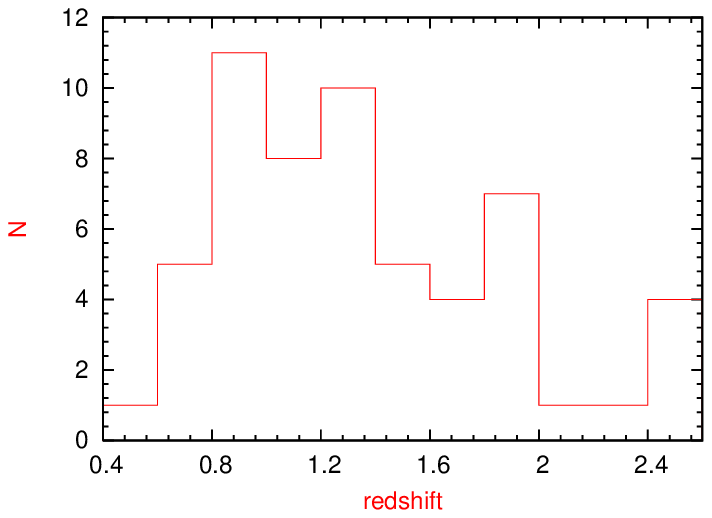} }
    \end{tabular}
   \caption{The photometric redshift and K band magnitude
   distribution for the $57$ EROs  counterparts (in Bo\"{o}tes field) in left
    and right panels respectively.} \label{histkzero}
 \end{center}
 \end{figure*}

Several selection criteria have been defined for EROs, including
$R-K \geq 6, R-K\geq 5.3, R-K \geq 5, I-K \geq 4$ with K-magnitude
upper limits from 18 to 21 mag. All these criteria are designed to
find evolved galaxies at high redshift (Cimatti et al 1999,
Abraham et al. 2004, Pozzetti \& Manucci 2000). The limit of $R-K =5$ is adopted in this study  as the definition of an ERO. Based on
this 57 ($19\% $)  FIRST radio sources are found  to have an ERO
as a counterparts with $R-K\geq 5$  and $16.5 \leq K \leq 20$.
This fraction includes sources that have identifications in five
bands (Bw, R, I, J, K), four bands (Bw, R, I, K) or less (i.e. three bands ((Bw, R, K); (R, I, K) or two bands (R, K)). As a comparison with ERO samples selected based on the criteria $(I-K)
\geq 4$, I found 41 EROs and  33 of the $(R-K) \geq 5 $ EROs have
$(I-K) \geq 4$. The R and I bands based selection criteria for
EROs are not equivalent. Figure \ref{rkcolor}  shows the $R-K$ vs.
$K$ (left panel)  and  $I-K$ vs $K$ (right panel) colour-magnitude
diagram respectively  for FIRST radio sources with IDs in both R,
K bands (305). This figure shows that there are no stars  (empty
diamonds) that have a colour redder than $R-K \geq 5$. 30/57 sources
appear to have a magnitude $ \leq 18$ and 27/57 with a magnitude $
> 18$. Figure \ref{histkzero} shows the $K$ band magnitude distribution (left
panel) and photometric redshift distribution (right panel) of all
57 EROs. The median $K$ band magnitude of these 57 EROs is K$=
19.27 \pm 0.72$ and  is equal to K$=17.97 \pm 0.71$ when sources
with $z>2$ excluded.

\subsection{ Colour-colour separation}

Since EROs are composed of both passively evolving red galaxies
and dusty starburst, colours are one means of breaking the
degeneracy between these two groups. The SED of an elliptical
galaxy at $z\sim 1$ drops off sharply at the $4000 A^{\circ}$
between \textit{R} and \textit{K} bands, while the SED of a dusty starburst declines
more gradually because of reddening. Therefore observations
between the \textit{R} and \textit{K} bands, such as \textit{I, Z, J or H} can be used to
measure the sharpness of the spectral type break and thus
discriminate between these two scenarios (Pozzetti \& Mannucci
(2000)). The ($J-K$) separates EROs at $z>1$ to bluer early type
galaxies and to redder  dusty EROs. As there is  no J
band for all FIRST radio sources IDs; I used only  the EROs
counterparts of FIRST radio sources identified in the second strip
($33 ^{\circ}\leq\delta<34^{\circ}$) where the J band data are
available from FLAMEX. There are 30 EROs in this strip where 25/30
were detected in J band.

\begin{figure}
\begin{center}
 \includegraphics[width=83mm]{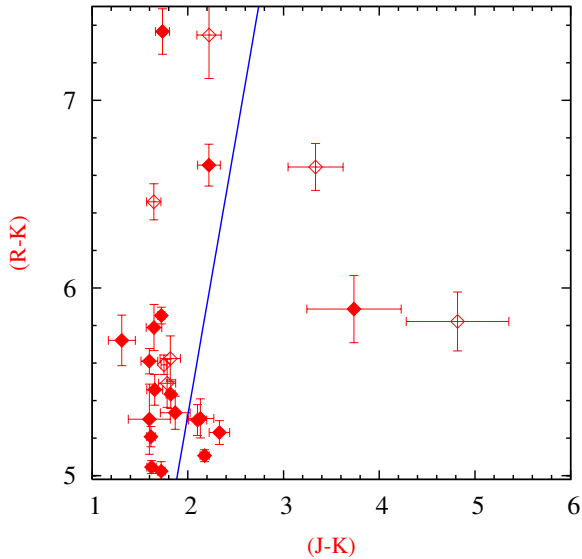}
\caption{\textit{R-K} against \textit{J-K} Colour-colour diagram for the 25
counterparts of FIRST  radio sources that have \textit{J} band (provided by
FLAMEX) with \textit{R-K} $>5$. The solid line represents the boundary
between dusty starburst and evolved, passive magnitude EROs
proposed by Pozzetti \& Mannucci (2000); the dusty galaxies
  should lie to the right of the line (redder (\textit{J-K}) colours) with the evolved
  systems on the left.}
\label{pozetti}
\end{center}
\end{figure}

Figure \ref{pozetti} shows $R-K$ versus $J-K$ colour-colour scheme
introduced by Pozzetti \& Mannucci (2000), where elliptical
galaxies lie in the left  (bluer $J-K$), and dusty starburst lie
to the right (redder $J-K$ colour). I find 7/25 ($18\%$)  EROs
selected in the dusty starburst side of the indicator and 18/25
($72\%$) appear elliptical. The colour-colour scheme
introduced by Bergstrom \& Wiklind (2004) using the ($R-J$) colour
vs. ($J-K$) colour is also used. There is a total agreement with Pozzetti \&
Mannucci (2000) scheme. But no reasonable agreement was found between
the classification using the SED template fitting method and the
optical/near-infrared colours of the sample. There are 11 sources
(with $\chi^2<2.7$, $90\%$ confidence limit ) classified as burst
using \textit{Hyperz} and classified as elliptical in the colour-colour
diagram (see following section).

\subsection{Photometric classification of EROs}

As a second way to classify an EROs sample to whether passively evolving red
galaxies, or dusty starburst galaxies, I use the same method described by
Smail et al. (2002). The \textit{Hyperz} code was used to find the best-fit ($\chi^{2}$ minimum);
empirical galaxy template by comparing the EROs colours to two different
families of SEDs and sought the best solution: dusty,  young starburst (or
AGN) or almost dust free evolved systems. All models in \textit{Hyperz} are built with
GISSEL 98 (which is an update of the spectral synthesis models described by
Bruzual \& Charlot (1993)) with Miller \& Scalo  (1979) initial mass function,
solar metalicity and different star formation rate (SFR). For the dusty
starburst I adopted a continuous star formation model and the reddening
($A_v$) ranges between 1 and 4 (Cimatti et al 2002). While for the evolved
galaxies I used a model SED represented by exponentially decaying star
formation rate with e-folding time of 1 Gyr, the reddening for this model is
required to be $A_v < 0.5$. For these model fits I adopted the Calzetti et
al. (2000) reddening law. Since I do not have J band data  for all the sources,
the  \textit{Hyperz} was run  for  two samples: the first sample included only EROs
counterparts of FIRST that have J band data (provided by FLAMEX; $33^{\circ}
\leq \delta<34^{\circ}$), there are 25/57 sources with J data and 20/25
detected in 5 bands (\textit{Bw, R, I, J, K}). The second sample contains 27/57 that do
not have J band data (third and fourth strips; $34^{\circ}
\leq\delta<36^{\circ}$)  and 17/27 identified in four bands (Bw, R, I, K). The
photometric redshift predictions of three fits (Ell, Burst, all BC SED
templates (E, Burst, SO, Sa, Sb..)) are shown in table \ref{ero2}. One notes that,  the
best-fit elliptical template has a lower photometric redshift than the
best-fit starburst template. The mean photometric redshift is $z= 1.35$ and
the rms scatter in the photometric redshift predictions for these three fits
is $\sigma_z<0.51$; the photometric redshifts are thus  quite robust and show
clearly that these EROs are $z \geq 0.8$ galaxies (except two sources one at $z
= 0.605$ classified as a dusty source using the colour-colour plot and the
second one at $z =0.55$ classified also as dusty source according to the
template fitting with a high probability). When all the SED templates are
allowed; 14/57 sources cannot provide an acceptable fit giving $\chi^2>2.7$;
11/57 are classified as elliptical and 31/57 as starburst. When the elliptical
option is chosen 28/57 has $\chi^2<2.7$ ($90\%$ confidence limit) and 22/57
has $\chi^2>2.7$ when  the starburst option is chosen.  It is important to note that broad
band photometry alone cannot  provide a reliable information on the spectral
type due to the degeneracy  between age, metalicity and reddening (see  Fig 1
in Bolzonella, Miralles \&  Pello 2000). Nevertheless it provides a rough SED
classification at two extreme: a given object has "blue " or "red" continuum
at a given $z$. The early type galaxies
can be matched with elliptical or burst templates. These SEDs are nearly
identical (Bolzonella, Miralles \&  Pello, 2000) and  the best-fit  \textit{Hyperz}
'starburst' are evolved results of instantaneous burst and not  really dusty
star-forming galaxies. Table \ref{ero2} presents the object name; 1.4 GHz radio flux
density, colours ($R-K$ and $J-K$), the $K$ band magnitude,
Photometric redshift and best fitting SED ($ \chi^{2} $) (either,
"starburst", "elliptical") of all EROs counterparts of FIRST radio
sources.  One of the dusty sources is assigned redshift $z<0.8$
($z=0.605$) lower than that expected for EROs.

\section{conclusions}

In this paper I have discussed the nature of the faint radio population at the
mJy level by making use of a sample of $\sim 700$ objects drawn from the joint
use of the FIRST  survey and the NDWFS/FLAMEX surveys.  

I find that the population of faint radio sources is mainly
dominated by early type galaxies, with radio to optical ratio between $10^2$
and $10^6$ and a very red colour $ Bw-R \, \sim 4.6$ confirming the results
found by Magliocchetti et al. (2002) who studied the optical counterparts of
FIRST radio sources in the APM survey to a limiting magnitude of B$\sim21$ and
$R\sim 20$. 

I  find 13 DRGs counterparts to FIRST radio sources in Cetus field with
$J-K>2.3$ to $K<20$. Obviously deep optical imaging and  spectroscopy  is
required to investigate the nature of these DRGs.

Taking advantage of the infrared data provided  by FLAMEX survey J and K
bands, I find 57 EROs counterparts to FIRST radio sources (in Bo\"{o}tes
field) with $R-K \geq 5$. I make use of a $J-K$ vs. optical infrared
colour-colour diagram (Pozzetti \& Mnuucci 2000) to separate EROs into
passively evolving and dusty star-forming galaxies  finding  that there is
$18/25$ ($72\%$) of EROs  are early type  galaxies with $z=0.6-2$ and $7/25$
($28\%$) dusty star forming galaxies. $3/7$ EROs from this latter group are DRGs with $ J - K
> 2.3 $.

Using the NDWFS/FLAMEX deep \textit{Bw R I J K} imaging I photometrically classify the EROs
counterparts to FIRST radio sources into dusty or evolved  based on simple SED
models.  The latter one cannot really provide a reliable information about the
spectral type due to the degeneracy between age, metalicity, but provides rough
SED classification at two extremes  either "blue" or "red" (Bolzonella,
Miralles \&  Pello, 2000).

Finally, Spectroscopic follow up of the faint radio population and their
surrounding EROs (and DRGs), deeper optical and infrared imaging is required
in order to establish colours for the faint radio population  and explore
their morphology in more detail. Deeper optical/infrared imaging will also
allow  to select deeper and more complete sample of EROs and to determine the
relationship  between the radio galaxies and the EROs allowing for a more
detailed  study of the faint radio population environment. It is proposed to follow up
the southern field (Cetus) spectroscopically using the Southern  African Large
Telescope.

\acknowledgements

 I would like to thank the anonymous referee for helpful comments and
 suggestion which improved the paper. I would like to thank  Dr Gonzalez
 Antony for answering all the questions concerning the  FLAMEX surveys.

I also would like to thank South African Square Kilometre Array project for
supporting and funding my Ph.D studies.

This work makes  use of images data products provided by the NOAO Deep 
Wide-Field Survey (Jannuzi and Dey $1999$), which is supported by
the National Optical Astronomy Observatory (NOAO). NOAO is
operated by AURA, Inc., under a cooperative agreement with the
National Science  Foundation.

 This publication makes use of data products  from the FLAMEX survey. FLAMEX
 was designed and constructed by the  infrared instrumentation group (PI:
 R. Elston) at the University of Florida, Department of Astronomy, with
 support from NSF grant AST97-31180 and Kitt Peak National Observatory.

\begin{table*}
\begin{minipage}{170mm}
\caption{Photometric redshift results for ERO sample. Cols (1)
Object name, colas(2) flux density of each radio source and (3)and
(4) colors; colas(5)-(6): Best-fit photometric redshift,
$X^{2}_{\nu}$ for a Gissel 98 elliptical galaxy; colas(7)-(8):
Best-fit photometric redshift, $X^{2}_{\nu}$ for a Gissel 98
starburst model;  colas(9)-(11): Best-fit photometric redshift,
$X^{2}_{\nu}$ when all SED template included (E, burst, SO, Sa,
Sb, Sc, Sd, Im) for a Gissel 98 elliptical galaxy.} \label{ero2}

\begin{tabular}{cccc|cc|cc|cc|c}
\hline
 \hline
   &   &   &   &    \multicolumn{2}{c}{All BC templates}  &   \multicolumn{2}{c}{Starburst} &   \multicolumn{2}{c}{Elliptical}  \\
\hline
Object name  & $S_{1.4} (mJy)$ & $R-K$  & $J-K$ &  $z$ & $\chi^{2}_{\nu}$   & $z$ & $\chi^{2}_{\nu}$   & $z$ & $\chi^{2}_{\nu}$ & SED  \\
(1)                   & (2)    & (3)       &  (4)    &  (5)    &    (6)  & (7) & (8)     & (9) & (10) & (11) \\
\hline
& & & & & & & & & & \\
NDWFS\_J142439.6+3349 &   2.71 &   5.43 &   1.82   &  1.075 &    0.26    &1.220 &    0.46  & 1.050 &    0.23 &           E \\
NDWFS\_J142449.3+3343 &   2.02 &   5.89 &   3.73  &  2.435 &    0.61   &2.445 &    4.63  & 1.395 &    3.86 &           E \\
NDWFS\_J142525.2+3345 &   4.66 &   5.02 &   1.72  &  1.345 &    0.20   &1.345 &    0.21  & 1.135 &    2.38 &           Burst \\
NDWFS\_J142611.0+3339 &   1.70 &   5.30 &   2.10  &  0.880 &    0.36  &1.215 &    1.56  &  1.030 &    1.64 &           SO \\
NDWFS\_J142616.8+3309 &   1.09 &   5.46 &   1.65  &  1.800 &    0.85   &1.800 &    0.85  & 1.255 &    5.97 &           Burst \\
NDWFS\_J142643.3+3351 &   2.86 &   5.79 &   1.65  &  1.555 &    0.89   &1.555 &    0.89  & 1.095 &    3.11 &           Burst \\
NDWFS\_J142755.9+3321 &  16.74 &   5.30 &   1.40  &  1.845 &    1.63   &1.780 &    8.58  & 1.845 &    1.63 &           E \\
NDWFS\_J142843.0+3326 &   5.97 &   5.33 &   1.87  &  1.770 &    6.69   &1.490 &   34.36  & 1.770 &    6.69 &           E \\
NDWFS\_J142843.4+3355 &   2.41 &   5.82 &   4.82  &  2.945 &    3.24   &2.945 &    3.24  & 2.495 &   18.38 &           Burst \\
NDWFS\_J142910.3+3358 &   2.24 &   5.72 &   1.31  &  1.740 &    2.14   &1.740 &    2.14  & 1.305 &    3.13 &           Burst \\
NDWFS\_J142915.2+3303 &   1.79 &   7.37 &   1.73  &  1.675 &   13.54   &1.675 &   13.54  & 1.095 &   67.44 &           Burst \\
NDWFS\_J142916.1+3355 &   2.07 &   5.04 &   1.62  &  1.125 &    1.52   &1.125 &    1.52  & 0.935 &    3.16 &           Burst \\
NDWFS\_J143004.7+3302 &   3.18 &   5.61 &   1.60  &  1.580 &    0.46   &1.580 &    0.46  & 1.095 &    3.58 &           Burst \\
NDWFS\_J143116.7+3354 &   1.69 &   5.62 &   1.82  &  0.960 &    4.08   &0.960 &    4.08  & 0.840 &    4.87 &           Burst \\
NDWFS\_J143209.1+3356 &   1.93 &   5.11 &   2.17  &  0.605 &    2.16   &0.885 &    4.90  & 0.800 &    4.90 &           E \\
NDWFS\_J143226.7+3330 &   4.53 &   5.49 &   1.78  &  1.945 &    0.52   &1.945 &    0.52  & 1.095 &    4.49 &           Burst \\
NDWFS\_J143249.7+3316 &   1.32 &   5.23 &   2.33  &  1.005 &    6.91  & 1.005 &    6.91  & 0.840 &   12.61 &           Burst \\
NDWFS\_J143258.3+3315 &   0.81 &   5.85 &   1.72  &  1.480 &    0.46  & 1.480 &    0.46  & 1.010 &   32.56 &           Burst \\
NDWFS\_J143304.2+3334 &   2.95 &   5.59 &   1.75  &  1.595 &   99.42  & 1.595 &   99.42  & 1.395 &  174.09 &           Burst \\
NDWFS\_J143347.0+3353 &   1.19 &   7.34 &   2.22  &  1.700 &    0.00  & 1.700 &    0.00  & 1.395 &    9.54 &           Burst \\
NDWFS\_J143428.0+3311 &  23.23 &   5.30 &   2.13  &  1.820 &    9.84  & 1.490 &   30.71  & 1.820 &    9.84 &           E \\
NDWFS\_J143435.1+3343 &   5.62 &   5.21 &   1.61  &  1.295 &    0.36  & 1.295 &    0.36  & 1.000 &    2.36 &           Burst \\
NDWFS\_J143527.9+3311 &  39.60 &   6.65 &   2.22  &  1.650 &    1.10  & 1.530 &   16.12  & 1.395 &    7.84 &           E \\
NDWFS\_J143646.0+3345 &   3.22 &   6.64 &   3.33  &  2.545 &    0.86  & 2.545 &    0.86  & 1.395 &    8.13 &           Burst \\
NDWFS\_J143731.0+3340 &   2.64 &   6.46 &   1.64  &  1.650 &    1.20  &1.650 &    1.20  & 1.385 &   12.89 &           Burst \\
NDWFS\_J142517.2+3415 &  14.74 &   5.00 & -  &  1.580 &    8.18 &  1.580 &    8.18 &  1.100 &   14.74 &           E \\
NDWFS\_J142648.2+3458 &   4.46 &   5.70 & -  &  1.305 &    0.25 &  1.020 &    1.20 &  1.305 &    0.25 &          Burst  \\
NDWFS\_J142802.4+3437 &   1.20 &   6.49 & -  &  1.105 &    2.85 &  1.030 &    3.87 &  1.105 &    2.86 &           Burst \\
NDWFS\_J142850.6+3453 &   2.19 &   5.75 & -  &  1.015 &    0.11 &  0.840 &    3.29 &  1.015 &    0.11 &           Burst \\
NDWFS\_J142905.6+3449 &  31.42 &   5.04 & -  &  1.355 &    0.23 &  1.045 &    2.14 &  1.355 &    0.22 &           Burst \\
NDWFS\_J142917.6+3437 &   2.56 &   5.25 & -  &  0.840 &    3.40 &  0.805 &    5.33 &  0.840 &    3.39 &           Burst \\
NDWFS\_J142943.8+3434 &   1.89 &   6.00 & -&  1.300 &    0.03 &  1.035 &    1.53 &  1.300 &    0.03 &          Burst  \\
NDWFS\_J143248.7+3413 &   0.95 &   5.25 & -  &  0.555 &    0.05 &  0.735 &    0.72 &  0.555 &    0.05 &           Burst \\
NDWFS\_J143259.2+3406 &   1.14 &   6.56 & -  &  1.125 &    0.41 &  1.190 &    1.19 &  1.385 &    1.56 &           E \\
NDWFS\_J143308.0+3418 &   7.07 &   5.10 & -  &  0.805 &    2.45 &  0.795 &    6.11 &  0.805 &    2.44 &           Burst \\
NDWFS\_J143430.5+3427 &   1.50 &   5.81 & -  &  1.430 &    1.24 &  1.430 &    1.24 &  1.125 &    5.92 &           E \\
NDWFS\_J143506.4+3438 &   3.22 &   5.16 & -  &  0.935 &    8.67 &  0.840 &   13.65 &  0.935 &    8.67 &           Burst \\
NDWFS\_J143539.8+3443 &  23.22 &   5.74 & - &  0.780 &    0.21 &  0.830 &    6.64 &  0.895 &    0.22 &           E \\
NDWFS\_J142440.4+3511 &   1.06 &   6.11 & -  &  1.850 &    0.10 &  1.425 &   13.21 &  1.625 &   23.32 &           E \\
NDWFS\_J142618.6+3545 &   2.49 &   5.56 & -  &  1.860 &    0.02 &  1.070 &    1.14 &  1.860 &    0.02 &           Burst \\
NDWFS\_J142628.5+3527 &   1.79 &   5.39 & -  &  2.495 &    5.16 &  0.275 &    9.12 &  2.210 &    6.36 &           E \\
NDWFS\_J142639.3+3510 &   2.56 &   5.18 & -  &  0.850 &    0.10 &  0.870 &    0.00 &  1.070 &    0.20 &           E \\
NDWFS\_J142817.2+3509 &   5.46 &   6.31 & -  &  0.805 &    4.41 &  0.940 &    9.10 &  0.805 &    4.41 &           Burst \\
NDWFS\_J143042.5+3512 &   5.18 &   5.23 & -  &  2.495 &   17.63 &  0.260 &   20.76 &  2.415 &   18.45 &           E \\
NDWFS\_J143051.3+3543 &   3.41 &   5.51 & -  &  0.805 &    0.39 &  0.840 &    1.04 &  0.805 &    0.40 &           Burst \\
NDWFS\_J143108.3+3525 &   2.07 &   5.36 & -  &  1.335 &    0.07 &  1.020 &    0.32 &  1.335 &    0.07 &           Burst \\
NDWFS\_J143112.5+3535 &   9.64 &   6.08 & -  &  1.405 &    0.02 &  0.840 &    9.08 &  1.405 &    0.02 &           Burst \\
NDWFS\_J143134.5+3515 &  64.12 &   5.94 & -  &  0.800 &    0.00 &  1.010 &    1.28 &  0.800 &    0.00 &           Burst \\
NDWFS\_J143238.0+3530 &  17.86 &   5.20 & -  &  0.905 &    0.27 &  0.790 &    1.85 &  0.905 &    0.27 &           Burst \\
NDWFS\_J143246.7+3533 &   2.18 &   5.28 & -  &  1.825 &    0.05 &  0.605 &    6.61 &  1.825 &    0.05 &           Burst \\
NDWFS\_J143313.7+3539 &   0.90 &   6.04 & -  &  0.750 &    0.01 &  1.030 &    2.01 &  0.750 &    0.01 &           Burst \\
NDWFS\_J143414.5+3541 &   0.55 &   5.95 & -  &  1.240 &    0.00 &  1.240 &    0.00 &  1.495 &    0.83 &           E \\

\hline
\end{tabular}
\end{minipage}
\end{table*}

\end{document}